\documentclass{PoS}

\usepackage{bm}
\usepackage{amsmath}
\usepackage{cite}
\usepackage[utf8]{inputenc}

\DeclareMathAlphabet\mathbfcal{OMS}{cmsy}{b}{n}

\newcommand{\cm}{\ensuremath{\mathsf{cm}}}
\newcommand{\kk}{\ensuremath{{K\bar{K}}}}
\newcommand{\kkb}{\ensuremath{{K\!\bar{K}}}}
\newcommand{\pe}{\ensuremath{{\pi\eta}}}
\newcommand{\pep}{\ensuremath{{\pi\eta^\prime}}}

\title{Resonances in Coupled-Channel Scattering}

\ShortTitle{Resonances in coupled-channel scattering}

\author{\speaker{David J. Wilson}\\
        \\
       {\rm{(for the Hadron Spectrum Collaboration)}}\\
        \\
        Department of Applied Mathematics and Theoretical Physics,\\
        University of Cambridge, Wilberforce Road,\\
        Cambridge, CB3 0WA, UK.\\
        \\
        School of Mathematics, Trinity College Dublin,\\
        College Green, Dublin 2, Ireland.\\

        E-mail: \email{djwilson@maths.tcd.ie}}

\abstract{
Excited hadrons are seen as resonances in the scattering of lighter stable hadrons like $\pi$, $K$ and $\eta$. Many decay into multiple final states necessitating coupled-channel analyses. Recently it has become possible to obtain coupled-channel scattering amplitudes from lattice QCD. Using large diverse bases of operators it is possible to obtain reliable finite volume spectra at energies where multiple channels are open. Utilising the finite volume formalism proposed by L\"uscher and extended by several others, scattering amplitudes can be extracted from the finite volume spectra. Recent applications will be discussed where the energy dependence of scattering amplitudes is mapped out in several quantum numbers. These are then continued to complex energies to extract resonance poles and couplings.
}

\FullConference{34th annual International Symposium on Lattice Field Theory\\
		24-30 July 2016\\
		University of Southampton, UK \hfill {\rm DAMTP-2016-74}}

\begin{document}

\section{Introduction}
In hadron physics many of the excited states observed experimentally still elude rigorous understanding from the fundamental theory of Quantum Chromodynamics (QCD). These excited states are observed as short-lived resonant enhancements in the scattering cross-sections of lighter hadrons such as pions and kaons that are stable within QCD and thus survive long enough to trigger a particle detector.

In simple cases, such as the $\rho(770)$ resonance, only the $\pi\pi\to\pi\pi$ channel contributes, however for the majority of cases multiple final states are present. This includes old puzzles like the $a_0(980)$ and $f_0(980)$ seen around $K\bar{K}$ threshold, and new mysteries like the ``XYZ'' resonances seen in the charm energy region. The presence of multiple final states necessitates studying these resonances in a coupled-channel framework which has recently become possible using Lattice QCD.

In both experimental and theoretical analyses, the first step is to map out the scattering amplitude usually expressed as a function of the centre of mass energy, $E_\cm$. This can then be decomposed into partial waves with definite angular momentum, and analytically continued to complex energies to identify resonance poles and residues, which contain the masses, widths and relative coupling strengths to each scattering channel.

\section{Scattering in a finite volume}
In a finite cubic volume $L^3$ with periodic boundary conditions, the momentum $\vec{p}$ is quantised as $\vec{p}=\frac{2\pi}{L}\vec{n}$ where $\vec{n}$ can be any triplet of integers, and thus the infinite-volume continuum of hadron-hadron scattering energies is also quantised. In the absence of any interactions between the hadrons, the allowed values of energy are
\begin{align}
E=\left(m_1^2+\left(2\pi\vec{n}_1/L\right)^2\right)^\frac{1}{2} + \left(m_2^2+\left(2\pi\vec{n}_2/L\right)^2\right)^{\frac{1}{2}}
\end{align}
for hadrons with masses $m_1$ and $m_2$. If the finite volume contains interactions between the hadrons then the finite volume energies are shifted away from these values, which can be mapped on to the infinite volume scattering amplitudes as was first derived by L\"{u}scher and extended to consider non-zero total momentum, unequal masses, coupled-channels and scattering particles with intrinsic angular momentum~\cite{Luscher:1990ux,Luscher:1991cf, Rummukainen:1995vs, Feng:2004ua, He:2005ey, Christ:2005gi, Kim:2005gf, Bernard:2008ax, Bernard:2010fp, Leskovec:2012gb, Gockeler:2012yj, Guo:2012hv, Hansen:2012tf, Briceno:2012yi, Briceno:2014oea}. In the calculations presented, the quantisation condition applied is
\begin{align}
\mathrm{det}\left[\bm{1}+i\bm{\rho}(E_\cm)\cdot \mathbf{t}(E_\cm)\cdot\left(\bm{1}+i\mathbfcal{M}(E_\cm, L)\right)\right]=0
\label{eq_det}
\end{align}
where $\bm\rho$ is a diagonal matrix with $\rho_i=2k_i/E_\cm$, the phase space in scattering channel $i$, $\mathbf{t}$ is the scattering $t$-matrix which is related to the $S$-matrix by $\mathbf{S}=\mathbf{1}+2i\sqrt{\bm{\rho}}\cdot\mathbf{t}\cdot\sqrt{\bm{\rho}}$, and the matrix $\mathbfcal{M}$ contains sums of the generalised Zeta functions subduced into the relevant finite volume little groups, as defined in ref.~\cite{Dudek:2012gj}. The determinant is taken over channel and partial wave indices.

This quantisation condition is well-established and has been derived in a variety of methods. It is valid for general hadron-hadron scattering amplitudes. Several closely related methods such as a Hamiltonian and effective field theories in a finite volume also find the same quantisation condition\footnote{Up to negligibly small exponentially suppressed corrections.}, with a fixed form of the infinite volume interactions~\cite{Bernard:2010fp,Wu:2014vma}.

\section{Extracting Lattice QCD spectra and elastic scattering}
In lattice QCD we compute Euclidean correlation functions, $C_{ij}(t)=\left<0\left|\mathcal{O}_i(t)\mathcal{O}_j(0)^\dagger\right|0\right>$ that decay exponentially, and at large times are dominated by the ground state of the quantum numbers of the operator $\mathcal{O}^\dagger$. In order to obtain the finely spaced spectra required for scattering calculations the variational method can be applied~\cite{Michael:1985ne,Luscher:1990ck,Dudek:2007wv,Blossier:2009kd}. A matrix of correlation functions is computed and diagonalised using a generalised eigenvalue problem,
\begin{align}
C_{ij}(t)v_j^\mathfrak{n}=\lambda_\mathfrak{n}(t,t_0) C_{ij}(t_0)v_j^{\mathfrak{n}}
\end{align}
where the finite volume spectrum $\left\{ E_\mathfrak{n} \right\}$ is obtained from the leading $t$ dependence of $\lambda_\mathfrak{n}(t,t_0) \sim e^{-E_\mathfrak{n}(t-t_0)}$.

In order to extract a complete spectrum in a given energy region, it is necessary to use a sufficiently large basis of operators with a diverse range of constructions. The bases used in the calculations presented below comprise constructions like $\bar{\psi}\Gamma D...D \psi$ that overlap well onto ``$q\bar{q}$'' contributions to the spectrum, and constructions that resemble hadron-hadron states projected into definite momentum  $\sum_{\vec{p}_1,\vec{p}_2}\mathcal{C}(\vec{p}_1,\vec{p}_2)\Omega_{\mathfrak{n}_1}^\dagger(\vec{p}_{1}) \Omega_{\mathfrak{n}_2}^\dagger(\vec{p}_{2})$. We use variationally-optimal operators $\Omega_{\mathfrak{n}}^\dagger$ formed from the eigenvectors $v_j^\mathfrak{n}$ of single-hadron correlation matrices~\cite{Thomas:2011rh,Dudek:2012gj}.

To illustrate the methods we begin with ref.~\cite{Wilson:2015dqa} that considers the $\rho$ resonance in $I=1$, $J^P=1^-$ $\pi\pi$ scattering with $m_\pi=236$ MeV in a cubic volume with $L = 4.3$ fm and $32^3$ spatial points, also covered in ref.~\cite{Woss:2016lattice}. The \emph{distillation} method~\cite{Peardon:2009gh} is used to efficiently compute all the Wick contractions required by QCD. The $\rho$ resonance is a well-established benchmark scattering calculation, having been widely studied by several groups at a range of pion masses~\cite{Aoki:2007rd,Feng:2010es,Lang:2011mn,Aoki:2011yj,Pelissier:2012pi,Dudek:2012xn,Bali:2015gji,Guo:2016zos,Bulava:2016mks,Fu:2016itp}.

\begin{figure}
  
  \begin{center}
    \includegraphics[height=6cm]{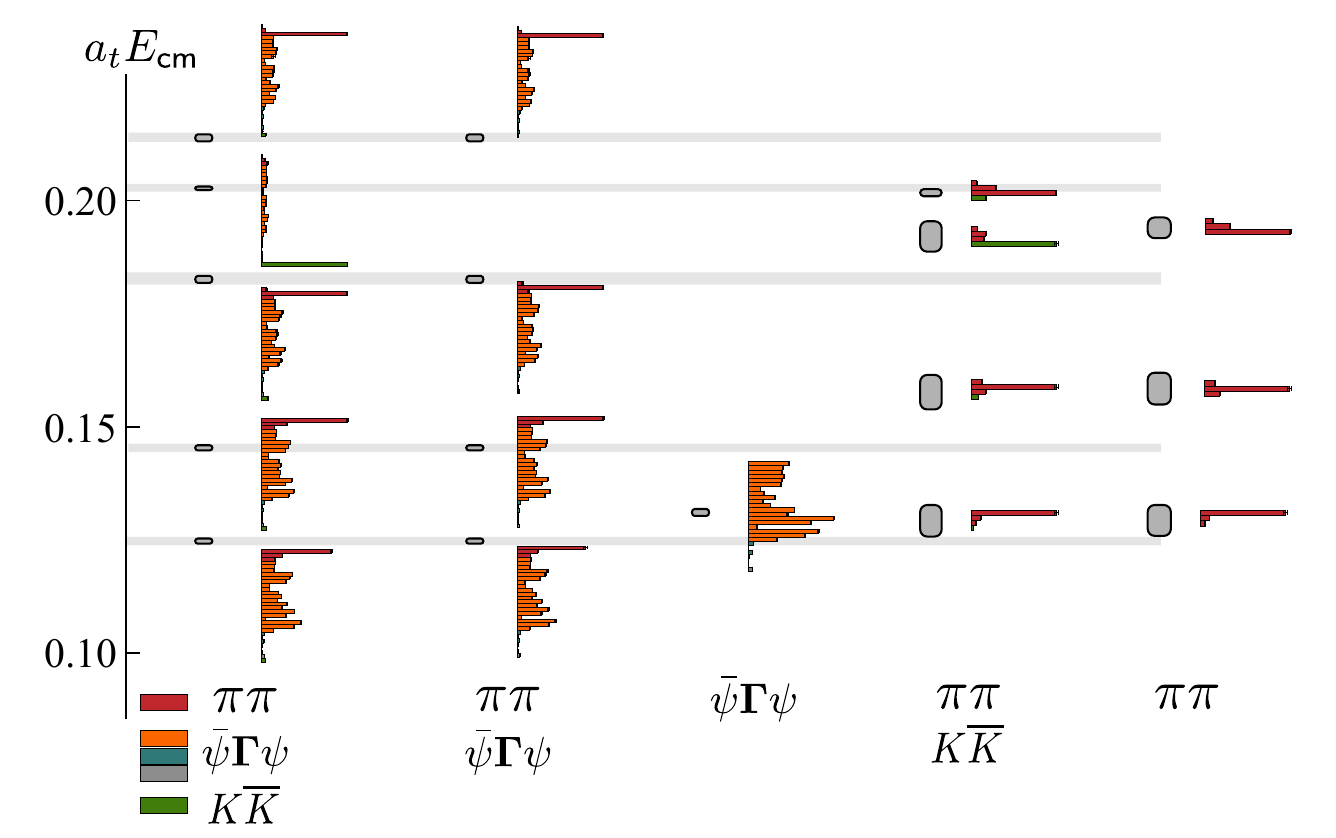}
    \caption{The spectra extracted from rest frame correlation matrices from ref.~\cite{Wilson:2015dqa}. The first column contains operators that resemble $\bar{q}q$, $\pi\pi$ and $K\bar K$ which yield five energy levels in this region. Using various subsets of these bases we find fewer states with larger uncertainties. The histograms show the contributions of each operator to the eigenvectors as suitably normalised overlap factors.}
  \label{fig_basis}
  \end{center}
  \vspace{-0.4cm}
\end{figure}

In scattering calculations, it is important to use a basis capable of interpolating every finite volume energy level present in a given energy region. Figure~\ref{fig_basis} shows the spectra that are extracted from five bases, the leftmost column of states uses the largest basis. Removing various classes of operator leads to fewer states, some with greater uncertainties, and some that are statistically precise but inaccurate. The horizontal bars represent the overlaps of different operators $\mathcal{O}^\dagger$ on to each lattice QCD eigenstate and there is considerable mixing between both the ``$q\bar{q}$'' and ``hadron-hadron'' contributions. Using only $\bar\psi\Gamma D...D\psi$ constructions yields just one state that is statistically incompatible with any of the states from the larger bases. In this example, it appears essential to have both the ``$q\bar{q}$'' and ``hadron-hadron''-like constructions in the operator basis.

\begin{figure}
  \vspace{-0.15cm}
  \begin{center}
    \includegraphics[width=0.96\textwidth]{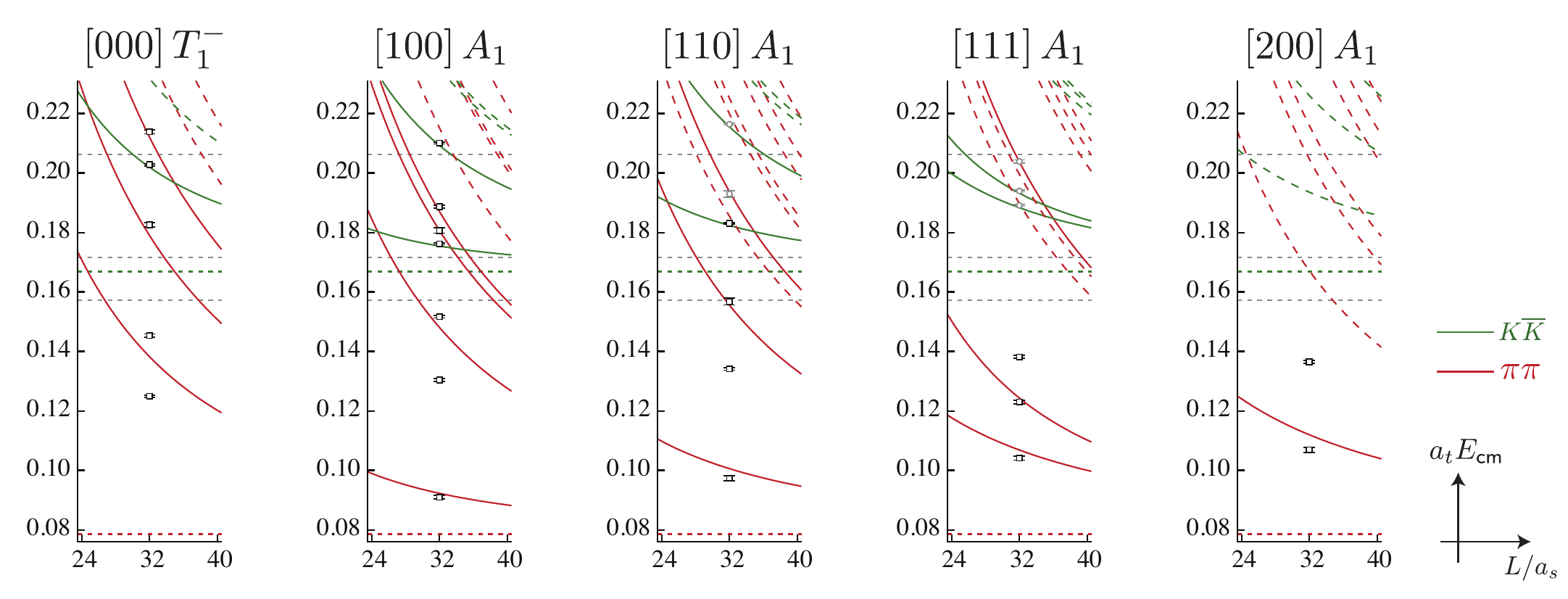}
    \caption{The finite volume spectra extracted in $I=1$ dominated by $J=1$ interactions in the energy region where the $\pi\pi$ and $K\bar{K}$ channels are open~\cite{Wilson:2015dqa}. The points are the energy levels, the solid lines are the non-interacting energies for meson-meson operators that were included in the basis; the dashed lines show non-interacting levels where operators were not included. Horizontal dotted lines indicate channel thresholds. The overall momentum is labelled with $\vec{d}=(i,j,k)=[ijk]$.}
    \label{fig_spec_rho}
  \end{center}
  \vspace{-0.4cm}
\end{figure}

Working in the rest frame, we find two levels in the single-channel region. In a finite volume, one efficient way of increasing the number of energy levels is to compute correlation matrices with an overall momentum $\vec{d}=\vec{n}_1+\vec{n}_2$, with respect to the lattice. Considering 5 moving frames up to overall momentum $\vec{d}^2=4$, this volume yields 22 energy levels below $4\pi$ threshold that can be used in the determinant condition~Eq.~\ref{eq_det} to determine the scattering amplitude. A selection of these are shown in Fig.~\ref{fig_spec_rho}.

\begin{figure}[!pth]
  \begin{center}
    \includegraphics[height=7.2cm]{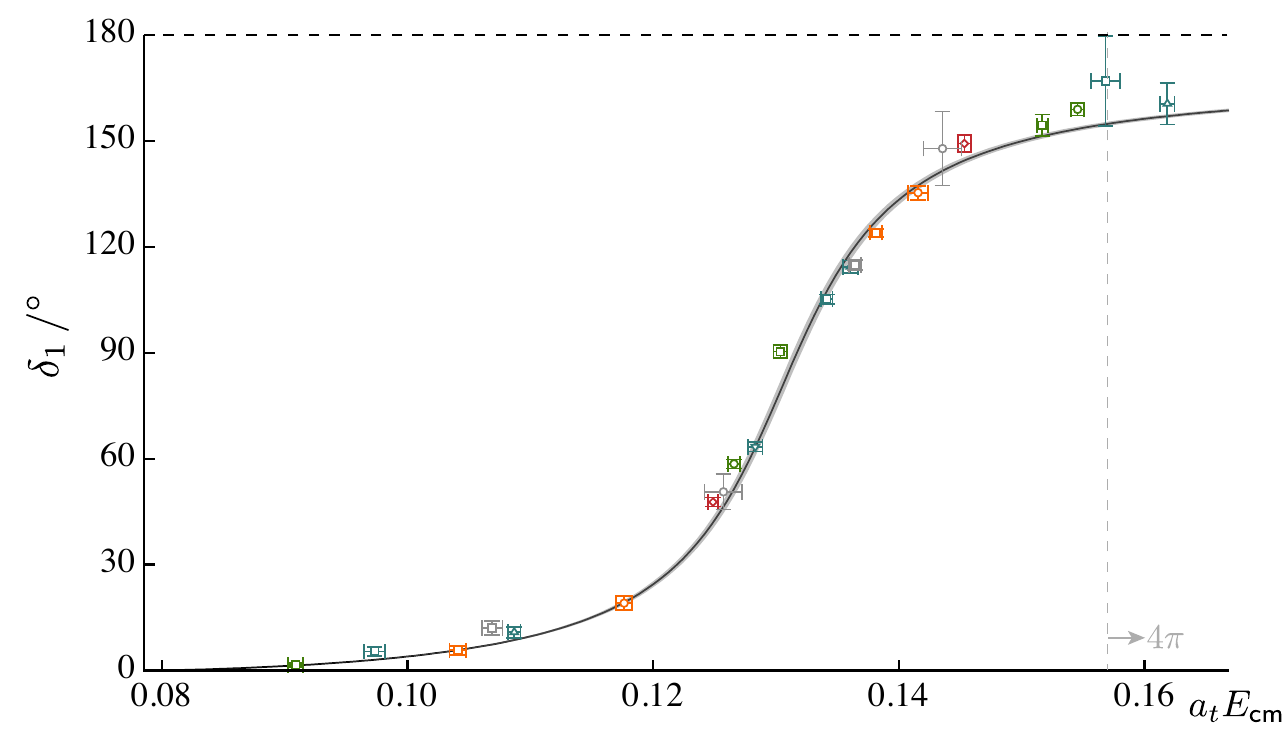}
    \caption{The elastic $\pi\pi$ phase shift in $I=1$, $J^P=1^-$ using the energy levels shown in Fig.~2 and five additional irreps that are not shown. Each discrete point corresponds to using an individual level to obtain a value of the phase shift from Eq.~2.2. The continuous curve corresponds to using a Breit-Wigner parameterisation in Eq.~2.2 minimised simultaneously using all 22 levels in the elastic region, from ref.~\cite{Wilson:2015dqa}.}
    \label{fig_phase_rho}
  \end{center}
\end{figure}

\begin{figure}[!pbh]
  \begin{center}
    \includegraphics[width=0.99\textwidth]{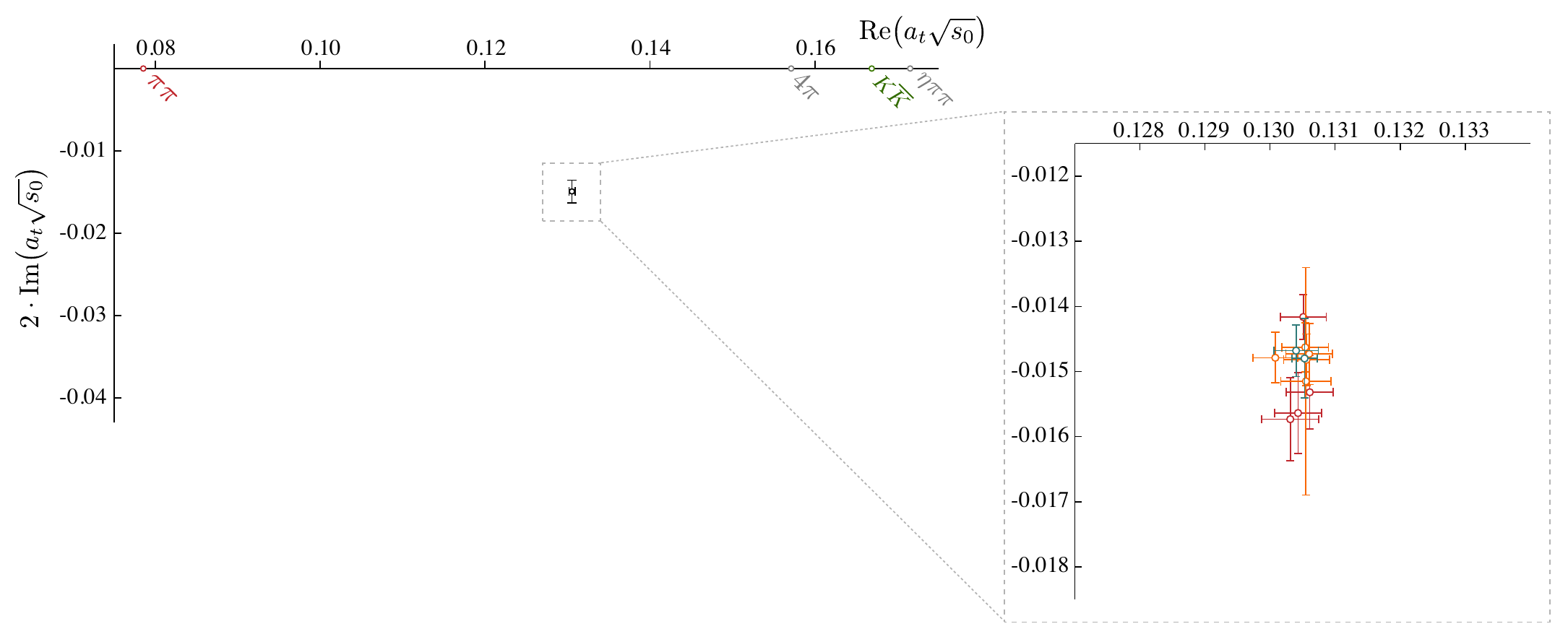}
    \caption{The position of the $t$-matrix pole of the form $t\sim \frac{c^2_{\pi\pi}}{s_0-s}$, where $s=E_\cm^2$, on the ``unphysical'' Riemann sheet which corresponds to taking the negative sign of the imaginary part of the scattering momentum. The real and imaginary parts in $\sqrt{s_0}$ correspond to the mass and width of a $\rho$ resonance, as computed in ref.~\cite{Wilson:2015dqa}. The zoomed panel shows the result of using many parameterisations, where very good agreement in both the real and imaginary parts is observed.}
    \label{fig_poles_rho}
  \end{center}
\end{figure}

In elastic scattering, the $t$-matrix can be written as $t=\frac{1}{\rho}\sin\delta e^{i\delta}$ and inserted into Eq.~\ref{eq_det} to relate each energy level to the scattering phase shift\footnote{Up to higher partial wave contributions, this is not discussed further here, but details are given in the references.} $\delta$. The discrete points shown in Fig.~\ref{fig_phase_rho} are obtained, showing a rapid variation of phase passing through $90^\circ$, indicating the presence of a nearby resonance pole, shown in Fig.~\ref{fig_poles_rho}.

A more generally applicable method to extract infinite volume scattering information is to parameterise the energy dependence of the $t$-matrix in Eq.~\ref{eq_det} to produce the whole spectrum at once. The parameters can then be minimised using a $\chi^2$ so that the lattice QCD spectra are well-described. One strength of this method is that it allows amplitudes with multiple unknowns at a given energy to be extracted, for example in coupled-channel scattering or if multiple partial waves contribute. Another strength is that these forms may be analytically continued to complex energies to extract resonance poles. One possible cause for concern is that a specific form may bias the result in some way, in order to circumvent this, a wide range of $t$-matrices can be used to infer that there is no dependence on this intermediate step. This is clear when considering the $t$-matrix poles in Fig.~\ref{fig_poles_rho} where statistically compatible results are found over many parameterisations.

\clearpage
\section{Coupled-channel scattering}
In order to compute the properties of resonances at higher energies where multiple channels contribute simultaneously, it is necessary to perform an analysis where the effects of coupling between the channels is included. For elastic scattering, the amplitude can be expressed as a single phase as a function of energy, $\delta(E_\cm)$ as was done above for the $\rho$ resonance. In coupled-channel scattering, the amplitude is a complex unitary matrix $\mathbf S$, with dimensions equal to the number of channels. In the scattering $S$-matrix for a two-channel system there are three unknowns as a function of energy\footnote{For each partial wave.}, so to simultaneously constrain these functions, multiple energy levels are required.

One useful form for plotting the elements of the $S$ or $t$-matrix is to use the phase and inelasticity, the simplest definition of this comes from the diagonal elements of the $S$-matrix where $S_{ii}=\eta e^{2i\delta_i}$. In terms of the $t$-matrix this reads,
\begin{align}
\mathbf{t}=\left(
\begin{matrix}
\frac{1}{2i\rho_1}\left(\eta e^{2i\delta_1}-1\right) &\quad \quad & \sqrt{1-\eta^2}\,\frac{1}{2\sqrt{\rho_1\rho_2}}\,e^{i\delta_1+i\delta_2} \\
\sqrt{1-\eta^2}\,\frac{1}{2\sqrt{\rho_1\rho_2}}\,e^{i\delta_1+i\delta_2} & \quad \quad & 
\frac{1}{2i\rho_2}\left(\eta e^{2i\delta_2}-1\right)
\end{matrix}
\right)
\label{eq_dde}
\end{align}
where $\eta$ is the inelasticity, and $\delta_i$ is the phase shift in channel $i$. These are just real functions of $E_\cm$. When only one channel is kinematically open, $\eta=1$ and only the phase of the open channel is relevant. In the two-channel region, unitarity of the $S$-matrix requires $\eta\in[0,1]$ with smaller values indicating a larger coupling between the channels.

In order to make use of the constraint provided by all the energy levels in a given region, which typically do not coincide, some interpolating function is required. Unitarity fixes the imaginary part of the inverse $t$-matrix for real energies to
$\mathrm{Im}\,\mathbf{t}^{-1}=-\bm{\rho}\,,$
thus the scattering amplitude can be written in terms of some matrix $K$
\begin{align}
\mathbf{t}^{-1}=\mathbf{K}^{-1}-i\bm{\rho}\,,
\end{align}
that is real for real $E_\cm$. Any form can be used, however certain terms are known to exist, for example from dispersion relations one can identify logarithms that must be present\footnote{These logarithms also arise when considering meson loops, for example at next-to-leading order in chiral perturbation theory.} due to the simple phase space $\rho$. This is used to form the Chew-Mandelstam phase-space $I(E_{\cm}^2)$, presented in the appendices of ref.~\cite{Wilson:2015dqa}, and is widely used in infinite volume amplitude analyses due to it having better analytic continuation properties.

The finite volume spectra can then be utilised to constrain many forms of $K$ that contain a few parameters each, these parameters are then over-constrained and a minimisation procedure can be performed where the spectrum from Eq.~\ref{eq_det} is compared with that found from the variational analyses of lattice QCD correlation functions.

\begin{figure}
  \begin{center}
    \includegraphics[width=0.99\textwidth]{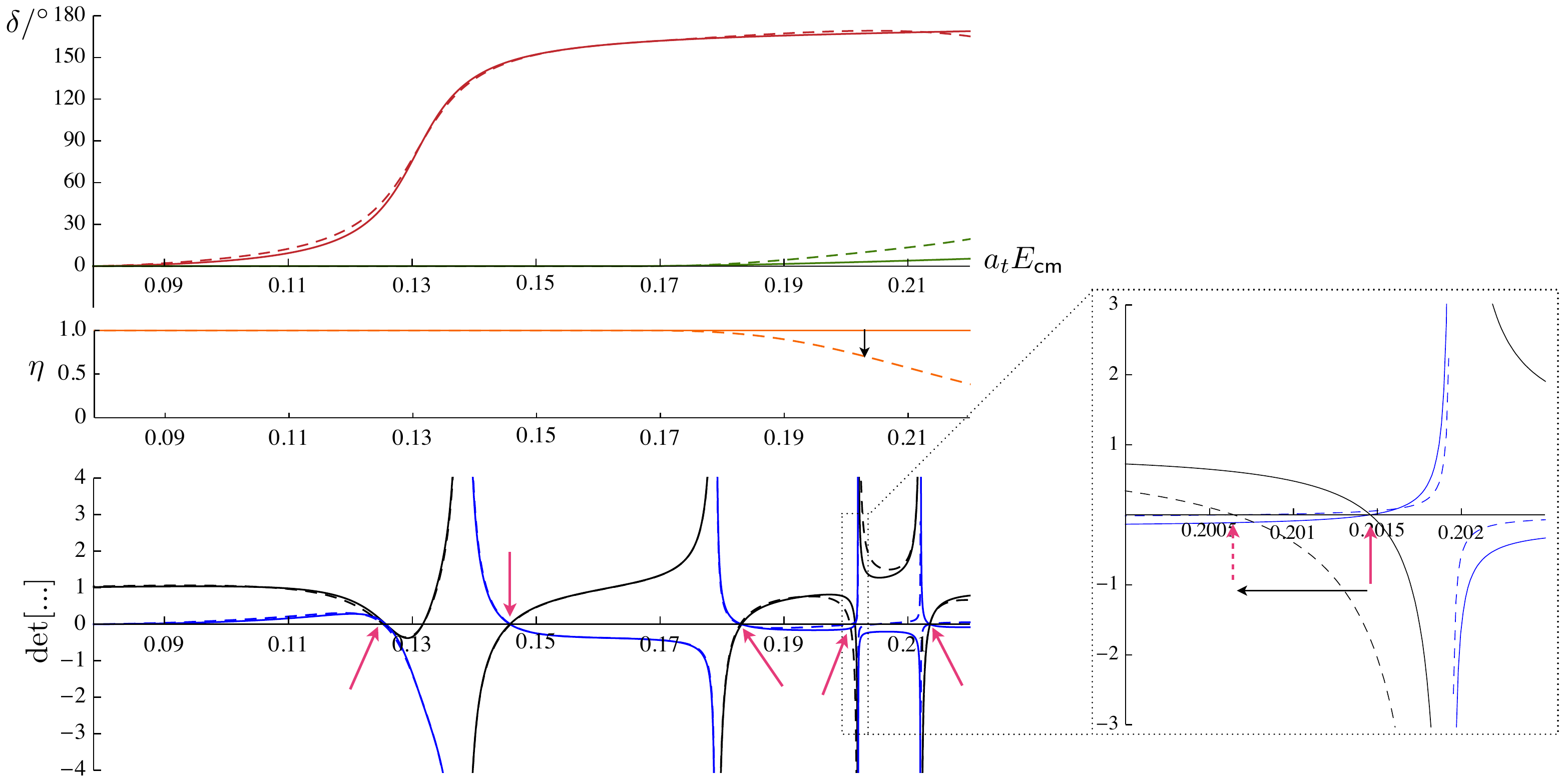}
    \caption{An illustration of the size of the effects to be probed when performing a coupled-channel analysis. The upper panel shows phase shifts $\delta_i$ and inelasticity $\eta$ (orange), for a $\pi\pi$ (red), $K\bar{K}$ (green) system. The real (black) and imaginary (blue) parts of the determinant are shown in the bottom panel. The solid curves are a $K$-matrix parameterisation from the $\rho$ calculation described in section 3. The dashed curves show the result of artificially increasing the off-diagonal elements of the $t$-matrix moving the inelasticity $\eta$ from 1. In the determinant, one of the solutions marked by a pink arrow, shows a shift of the order of $1\%$ due to this change in the amplitude, these details can be probed using lattice QCD calculations.}
    \label{fig_det}
  \end{center}
  \vspace{-0.5cm}
\end{figure}

In the simplifying limit that there are two decoupled channels, two independent finite-volume spectra are superimposed. An example of this is given by the solid curves in Fig.~\ref{fig_det}, for the case of the coupled $I=1$ $\pi\pi$, $K\bar K$ system in the $T_1^-$ irrep, extending the $\rho$ resonance described above into the coupled-channel region. Taking a coupled-channel $t$-matrix and artificially increasing the size of the off-diagonal elements, the dashed curves in Fig.~\ref{fig_det} are obtained. This illustrates the scale of the effects that must be probed in order to obtain coupled-channel scattering information. Differences in the energy levels of around $1\%$ are seen, which can be computed using lattice QCD.

\section{An $a_0$ resonance}
In the region around $K\bar K$ threshold in $J^P=0^+$, in both $I=0$ $\pi\pi$ and $I=1$ $\pi\eta$ scattering, puzzling effects are observed as the $K\bar K$ channel opens. In $I=1$, a range of phenomenological parameterisations working from experimental data produce a sharp peak in the $\pe\to\pe$ cross section at $K\bar K$, see for example ref.~\cite{Baru:2004xg}. The first lattice calculation was presented in ref.~\cite{Dudek:2016cru} and summarised in these proceedings~\cite{Dudek:2016lattice}.

\begin{figure}
  \begin{center}
    \includegraphics[width=0.7\textwidth]{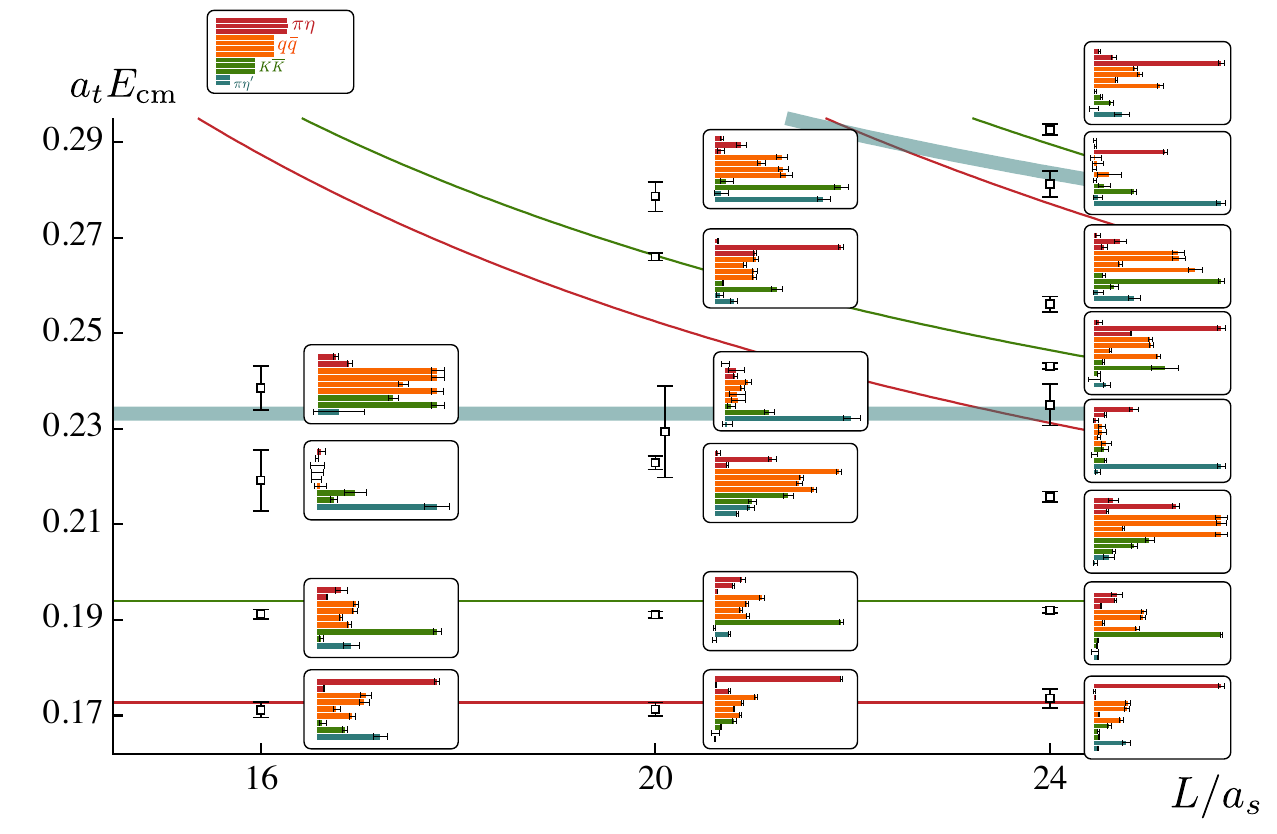}
    \caption{The spectra and operator overlaps obtained in the $A_1^+$ irrep with $I=1$, dominated by $J^P=0^+$ in the energy region where $\pi\eta$, $K\bar K$ and $\pi\eta^\prime$ channels are kinematically open. Many levels are significantly shifted from the non-interacting energies suggesting the presence of significant meson-meson interactions. The eigenvectors show a large amount of mixing between the operators used, as indicated by the horizontal bars~\cite{Dudek:2016cru}.}
     \label{fig_a0_spec_histo}
  \end{center}
\end{figure}

We compute a matrix of correlation functions using a large basis of operators with $\bar{q} q$-like and meson-meson-like constructions projected into definite lattice momenta, for each $\pe$, $\kk$ and $\pep$ combination that could possibly contribute in the energy region of interest. The energy levels extracted from the variational method, for zero overall momentum are shown in Fig.~\ref{fig_a0_spec_histo}. Levels shown in black, are seen to have large shifts away from where they would be expected in the absence of interactions, indicated by the solid curves. These large differences suggest non-trivial interactions are present. Another indicator is that significant contributions from multiple types of operators are seen in the eigenvectors for each level, shown by the horizontal bars in Fig.~\ref{fig_a0_spec_histo}. In order to quantify these effects in terms of the scattering matrix, the determinant condition Eq.~\ref{eq_det} must be utilised.

In the method outlined at the end of the previous section, first a $K$-matrix must be constructed to describe this system. One possible form is
\begin{align}
K_{ij}=\frac{g_i\,g_j}{m^2 - s}+\gamma_{ij}
\label{eq_kmat}
\end{align}
where $g_i$, $m$ and $\gamma_{ij}$ are real parameters that are minimised to describe the finite volume spectrum. This form is capable of efficiently producing a $t$-matrix pole, however there is no requirement this pole be close to the real axis. Many other forms are used, such as representing the inverse $K$-matrix as a polynomial
\begin{align}
K_{ij}^{-1}=\sum_{n=0}^N c_{ij}^{(n)}s^n
\end{align}
where $c_{ij}^{(n)}$ are real numbers and $N$ is a small integer. Other variations include fixing various parameters to zero and varying the real part of $I(s)$, either switching it off altogether, or changing the subtraction point. Many minimisations are performed with these $K$-matrices, and we select all those capable of describing the spectra up to some maximum $\chi^2/N_{\mathrm{dof}}$. 

\begin{figure}
  \begin{center}
    \includegraphics[width=1.0\textwidth]{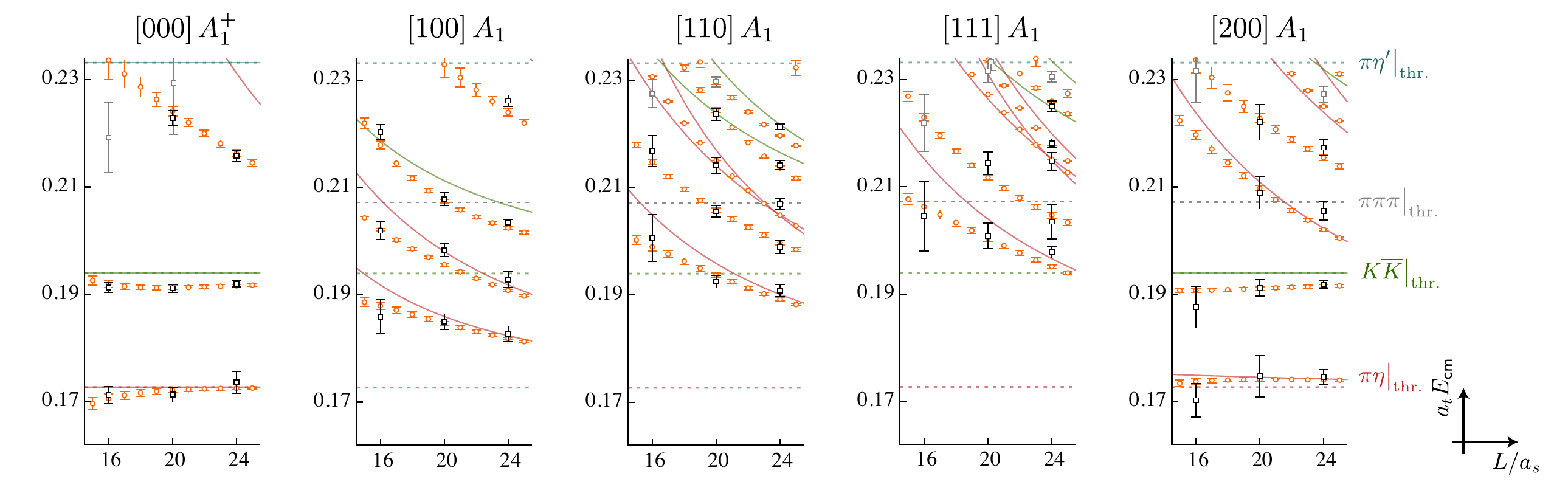}
    \caption{The spectra obtained in $a_0$ quantum numbers for 5 different moving frames from ref.~\cite{Dudek:2016cru}, shown in black. The orange points are the result of minimisation using a $t$-matrix in Eq.~2.2 defined by the $K$-matrix in Eq.~5.1 and a Chew-Mandelstam phase space, as described in the text.}
    \label{fig_a0_spec_fit}
  \end{center}
   \vspace{-0.5cm}
\end{figure}

\begin{figure}[tph]
\vspace{-0.1cm}
  \begin{center}
    \includegraphics[width=0.65\textwidth]{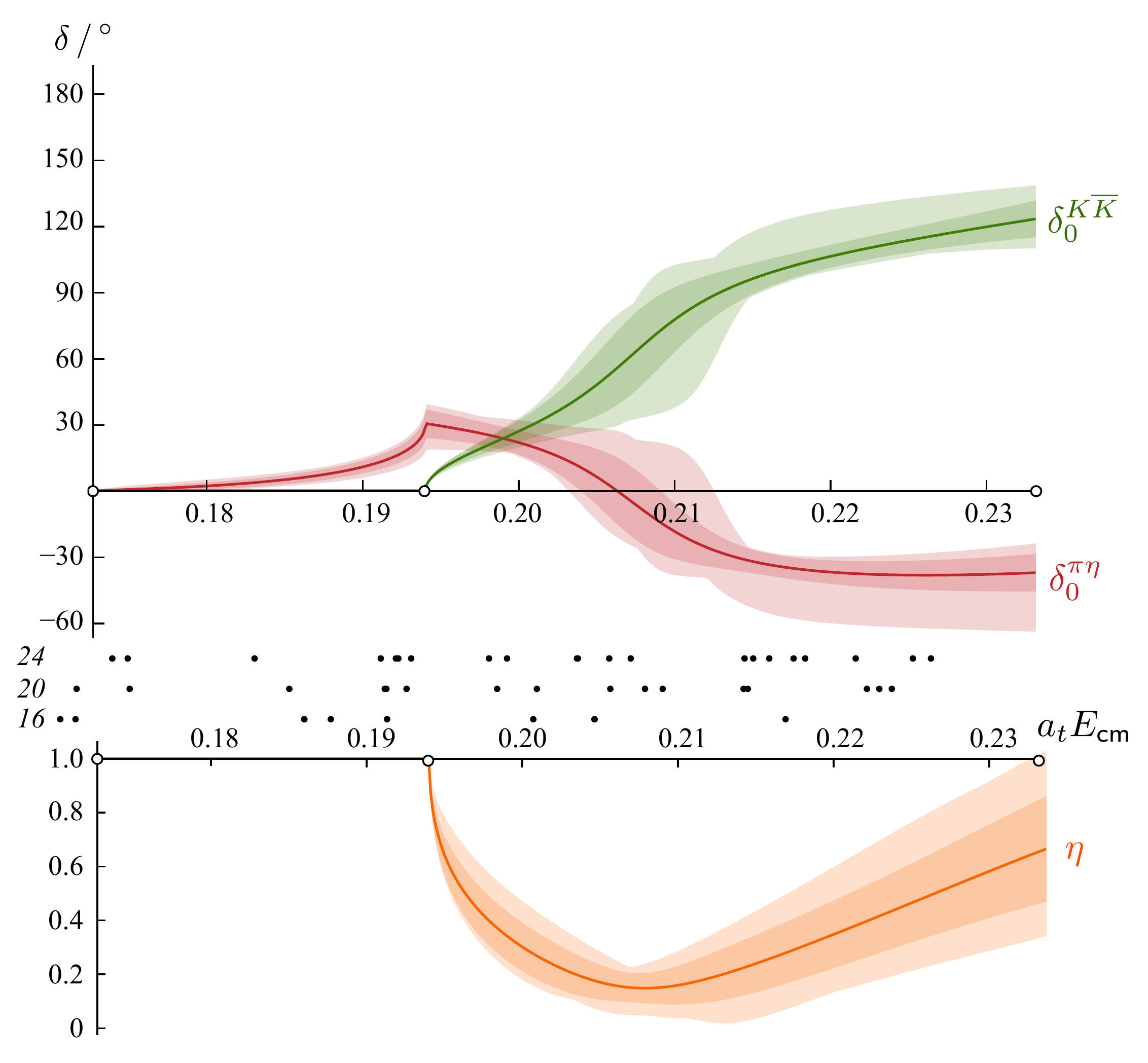}
    \caption{The phase shifts and inelasticity in the two-channel region in $a_0$ quantum numbers from ref.~\cite{Dudek:2016cru}. The central points show the positions of finite volume energy levels constraining the amplitudes. Rapid changes are seen just above $K\bar K$ threshold. The inelasticity $\eta$ deviates sharply from 1 and both channels undergo rapid variations in phase, due to a single nearby $a_0$ resonance pole.}
    \label{fig_a0_dde}
  \end{center}
  \vspace{-0.4cm}
\end{figure}

We begin with the two-channel region below $\pep$ threshold, where $\pe$ and $\kk$ are kinematically open. Using the energy levels given in Fig.~\ref{fig_a0_spec_histo} there is not much constraint, however this is remedied by using moving reference frames as described above. Considering boosts up to $\vec{d}=[200]$, 47 energy levels are obtained. These levels and the result of the minimisation using Eq.~\ref{eq_kmat} in Eq.~\ref{eq_det} are shown in Fig.~\ref{fig_a0_spec_fit}. This form reproduces the finite volume spectra well, and indeed this fit has $\chi^2/N_{\mathrm{dof}}=1.4$. In terms of more familiar quantities, we show the corresponding phase shifts $\delta_i$ and inelasticity $\eta$ in Fig.~\ref{fig_a0_dde}, as defined in Eq.~\ref{eq_dde} above.

In order to understand the resonance content of the scattering amplitudes, we analytically continue to complex energies to obtain the positions of the complex poles contained in these minimised $t$-matrices. In elastic scattering, there are two Riemann sheets connected along a square-root branch cut that runs along the real energy axis from threshold to $+\infty$. The first sheet is obtained by taking the root of $k_\cm$ where $\mathrm{Im}k_\cm > 0$, while the second sheet where resonance poles are found corresponds to taking the other sign. A similar picture emerges in coupled-channel scattering except now there are $2^n$ sheets to consider.

In each amplitude that successfully describes the finite volume spectra, we consistently find a pole of the form $t\sim\frac{c_i c_j}{s_0-s}$,  on the sheet corresponding to $\left(\mathrm{sign}(\mathrm{Im}\, k_{\pe}), \mathrm{sign}(\mathrm{Im}\, k_{\kk})\right)=(+,-)$, sometimes referred to as sheet $\mathsf{IV}$. This pole is very close to the boundary where this sheet connects to sheet $\mathsf{II}$ with signs $(-,+)$. This is interesting since the phase shifts undergo a flip as the pole moves from one sheet to the other, as is described in the appendices of ref.~\cite{Dudek:2016cru}. The amplitudes when plotted as $\rho_i\rho_j|t_{ij}|^2$ which is proportional to the cross-section, varies smoothly moving the pole from sheet $\mathsf{II}$ to sheet $\mathsf{IV}$, shown in Fig.~25 of ref.~\cite{Dudek:2016cru}.

In ref.~\cite{Dudek:2016cru} the possible positions of additional poles and the connection between the amplitude on the real axis and the poles in the complex plane was investigated. In order for a $t$-matrix pole to have a significant effect on the real axis it must be reasonably nearby. While many amplitudes contained additional poles none of these were found in close proximity to the physical scattering line, nor were they particularly well constrained, suggesting they cannot be very relevant to the physical scattering process.

\begin{figure}[bph]
  \begin{center}
    \begin{minipage}{0.6\textwidth}
    \includegraphics[width=1.0\textwidth]{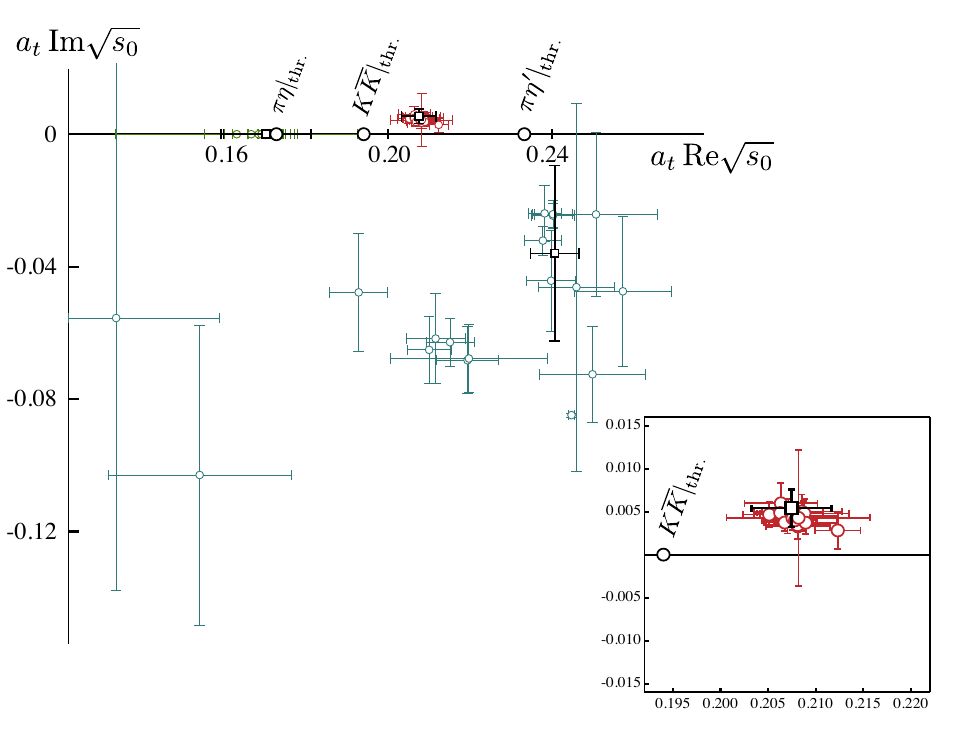}
    \end{minipage}
    \begin{minipage}{0.39\textwidth}
    \includegraphics[width=1.0\textwidth]{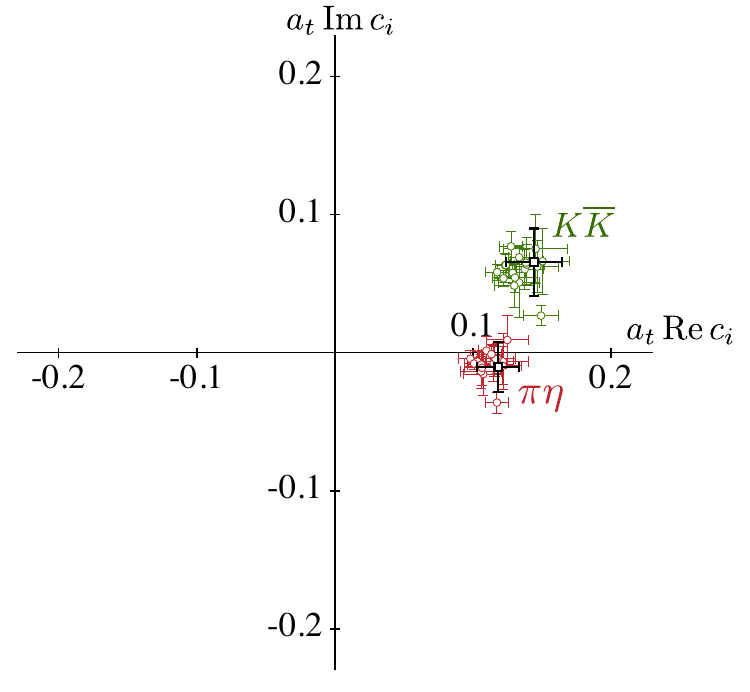}
    \end{minipage}
    \caption{Left: The poles in the $a_0$ amplitudes, where $t_{ij}\sim c_ic_j/(s_0-s)$ in the region of the pole. The well-determined sheet $\mathsf{IV}$ pole is shown in red and is present in every amplitude that successfully described the finite volume spectrum. Other poles were present in many of the amplitudes but only far from the real axis where physical scattering occurs. While not particularly well constrained, the distance of these poles indicates they are also much less relevant to physical scattering. The poles found when using the form in Eq.~5.1 are highlighted in black. Right: The factorised residue or ``couplings'' $c_i$ of the sheet $\mathsf{IV}$ pole in both the $\pe$ and $\kk$ channels, showing good agreement over many parameterisations.}
    \label{fig_a0_pole}
  \end{center}
\end{figure}

The factorised residues of the sheet~$\mathsf{IV}$ pole are obtained by considering each element of the $t$-matrix to extract $c_\pe$ and $c_\kkb$. These have a similar magnitude indicating the resonance has a significant coupling to both channels. These couplings are plotted in the right pane of Fig.~\ref{fig_a0_pole}.

An amplitude dominated by a single pole on one Riemann sheet is often associated with a significant molecular component~\cite{Weinberg:1965zz,Morgan:1992ge,Morgan:1993td}, in contrast to having nearby poles on two or more Riemann sheets. The behaviour of this amplitude as a function of the pion mass will bring additional insight into the origin and nature of this resonance pole.

\section{Other studies}

Using these methods, hadron-hadron scattering amplitudes with many quantum numbers can be computed. The first coupled-channel study using lattice QCD spectra was presented in refs.~\cite{Dudek:2014qha,Wilson:2014cna}, for the coupled $\pi K ,\, \eta K$ system, that also used a pion with $m_\pi=391$ MeV. The channels were found to be weakly coupled and a broadly rising $S$-wave $\pi K$ phase shift was extracted that was due to a broad $K_0^\star$ resonance pole. A virtual bound-state pole was also seen on an unphysical sheet below $\pi K$ threshold. Scattering amplitudes were also extracted for the $P$ and $D$-waves with a near-threshold $K^\star$ bound-state in $P$-wave and a narrow $K_2^\star$ resonance in $D$-wave.

\begin{figure}[!h]
\vspace{0.4cm}
  \begin{center}
    \includegraphics[width=0.46\textwidth]{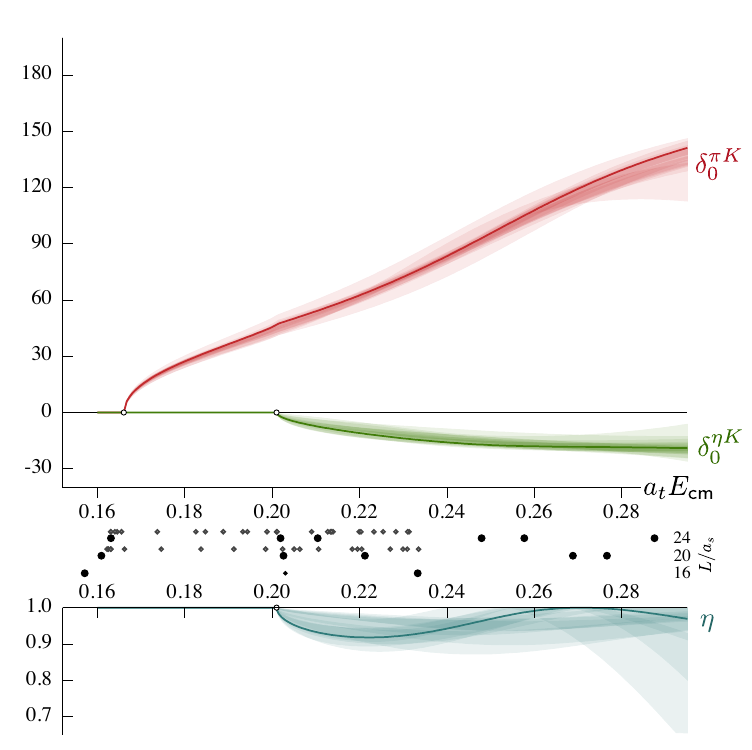}\includegraphics[width=0.53\textwidth]{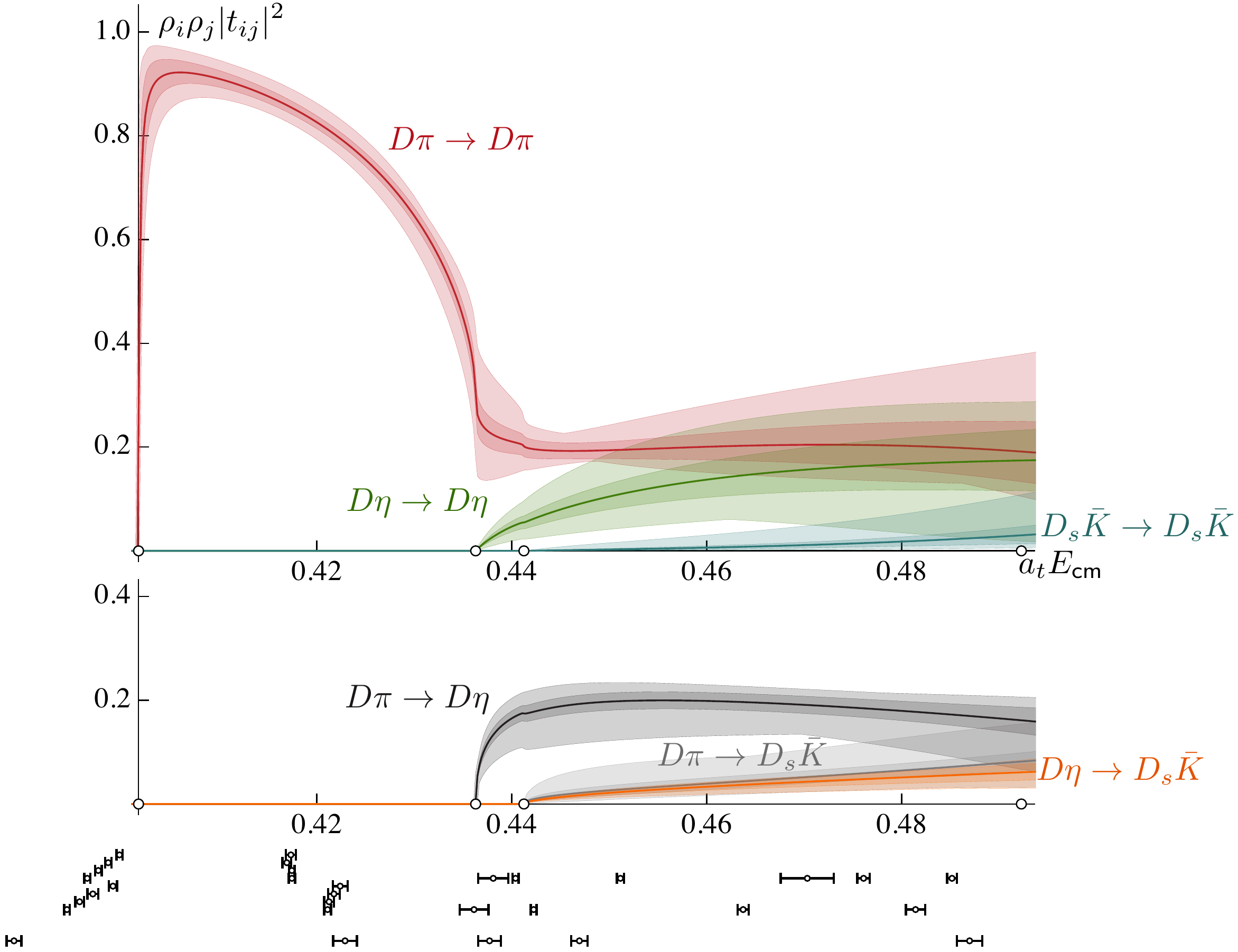}
    \caption{Left: The coupled $\pi K,\,\eta K$ system in $S$-wave, showing a broad enhancement in the $\pi K$ phase shift in red and small $\eta K$ phase shifts in green. The multiple bands include the variation over many parameterisations. Only small couplings between the channels were observed in this first study such that the inelasticity $\eta$ (blue) was found to be consistent with 1 over the energy region that was constrained. The central points show the locations of finite-volume energy levels. Right: The $D\pi,\,D\eta,\,D_s\bar{K}$ amplitudes in $S$-wave plotted as $\rho_i\rho_j|t_{ij}|^2$. The diagonal elements are shown in the top panel and off-diagonal elements are shown in the bottom panel. The points with errors show the positions of the lattice QCD energy levels. A broad feature is observed in the $D\pi\to D\pi$ amplitude due to a near-threshold bound-state with a large coupling to the $D\pi$ channel.}
    \label{fig_other_studies}
  \end{center}
\end{figure}

Recently these coupled channel methods have been applied to processes involving charm quarks, for $D\pi$ scattering~\cite{Moir:2016srx}, where only the elastic region had been considered previously~\cite{Mohler:2012na}. This coupled-channel study considered the three-channel $D\pi,\,D\eta,\,D_s\bar{K}$ system and extracted three partial waves. A broad feature was observed in $S$-wave due to a bound-state just below $D\pi$ threshold with a large coupling to $D\pi$. This produced a broad enhancement in the $D\pi\to D\pi$ amplitude, shown in the right panel of Fig.~\ref{fig_other_studies} plotted as $\rho_i\rho_j|t_{ij}|^2$ which is proportional to the cross-section. Only small effects were present in the other channels up to $D^\star\pi\pi$ threshold. In $D$-wave, a narrow $D_2^\star$ resonance was found with a dominant coupling to the $D\pi$ channel, and a deeply-bound $D^\star$ pole was computed in $P$-wave.

\clearpage
\section{Summary \& Outlook}
Recently, it has become possible to perform coupled-channel scattering calculations using lattice QCD, these have been applied at light and charm quark energies to two and three-channel hadron-hadron scattering. The $a_0$ resonance calculation was the first to extract a scattering amplitude from lattice QCD energy levels with strongly-coupled channels, indicated by the inelasticity $\eta$ approaching zero in Fig.~\ref{fig_a0_dde}. Although not presented here, scattering amplitudes were also extracted including the $\pep$ channel up to $\pi\pi\pi\eta$ threshold where Eq.~\ref{eq_det} is valid. In refs.~\cite{Dudek:2014qha,Wilson:2014cna}, the $\pi K ,\, \eta K$ coupled system was analysed, also at $m_\pi=391\, \mathrm{MeV}$ finding a broad $S$-wave resonance and weakly-coupled scattering channels. Recently, the three-channel $D\pi,\,D\eta,\,D_s\bar{K}$ system was also analysed where a $D^\star_0$ bound-state was found at $D\pi$ threshold~\cite{Moir:2016srx}.

Future studies will consider the $\pi\pi,\,K\bar K,\,\eta\eta$ system in $f_0$ quantum numbers, extending a recent elastic $\pi\pi$ study observing a $\sigma$ pole~\cite{Briceno:2016mjc}. Many systems with charm and charm-anticharm quark content have yet to be investigated. Calculations involving operators that resemble tetraquark objects are underway, and extensions of these methods to include electromagnetic probes have recently been completed~\cite{Briceno:2015dca,Briceno:2016kkp}, with further progress reported in these proceedings~\cite{Leskovec:2016lrm}. 

One challenge facing these methods when calculations are performed at lighter pion masses is that channels open with three or more hadrons, like $3\pi$ and $4\pi$ in meson channels and $N\pi\pi$ in baryon channels. Significant progress is being made to understand these three-body channels~\cite{Briceno:2016ffu,Hansen:2014eka,Hansen:2015zga}. Many interesting two-body channels remain, for example coupled-channel methods have yet to be applied to scattering of spinning particles, which are present low in the spectrum for systems like $N\pi$, $D^\star\pi$ or $J/\psi\pi$. 

These recent computations demonstrate that coupled-channel scattering amplitudes can be extracted using lattice QCD. The methods utilised show great promise for rigorously extracting properties of unstable states higher in the spectrum, and provide opportunities to make use of infinite volume amplitude analysis methods that compare directly with experiment.

\section*{Acknowledgements}
Contributions from colleagues in the Hadron Spectrum Collaboration are gratefully acknowledged. This work was supported by a grant from the Simons Foundation to the Hamilton Mathematics Institute at Trinity College Dublin, and the Isaac Newton Trust/University of Cambridge Early Career Support Scheme [RG74916].

\bibliographystyle{JHEP}
\bibliography{refs}

\end{document}